\documentclass[]{spie}  %>>> use for US letter paper
 \usepackage{makecell}
\usepackage{amsmath,amsfonts,amssymb}
\usepackage{graphicx}
\usepackage[colorlinks=true, allcolors=blue]{hyperref}
\usepackage{caption}
\usepackage{booktabs}
\usepackage{setspace}
\title{Design overview of the fiber feed for the Hobby-Eberly telescope's HRS}

\author[a]{Devika K Divakar}
\author[a]{Phillip MacQueen}
\author[a]{Joseph Strubhar}
\affil[a]{McDonald Observatory, The University of Texas at Austin, 2515 Speedway Blvd, Texas, 78712}
\authorinfo{Further author information: (Send correspondence to A.A.A.)\\A.A.A.: E-mail: devika.divakar@austin.utexas.edu}%\\  B.B.A.: E-mail: bba@cmp.com, Telephone: +33 (0)1 98 76 54 32}

\begin{document} 
\maketitle

\begin{abstract}
We present the redesign of the fiber feed for the High Resolution Spectrograph (HRS) at the Hobby–Eberly Telescope (HET). The upgrade incorporates a static atmospheric dispersion corrector (ADC) using Ohara i-line glasses (BAL15Y and S-FPL51Y), carefully selected for high internal transmission ($>$ 99\%), and optimized to improve throughput and image quality across the 360 - 1000 nm band. The ADC consists of two identical Amici prisms, fixed at an orientation optimized for the HET’s nominal zenith angle (35\textdegree), correcting dispersion over the HET zenith range of 26.5\textdegree - 43.5\textdegree. Relay optics were optimized to improve blue-end sensitivity and maintain substantially sub-fiber-core RMS spot radii across the full field of view. Simulations, including atmospheric dispersion modeling in ZEMAX show residual dispersion $\le$ 0.42" considering the entire range of wavelengths and zenith distances, with transmission efficiency of 91 - 94\%. We also discuss how the mechanical design integrates all optical elements, including the ADC, in a rigid, modular input head assembly mounted in the HET Prime Focus Instrument Package (PFIP). This optimized fiber feed enhances coupling efficiency, improves S/N in the blue, and enables higher radial velocity precision - maintaining HET - HRS as a leading facility for high-resolution spectroscopy.
\end{abstract}


\section{INTRODUCTION}
\label{sec:intro}  % \label{} allows reference to this section
\begin{figure}[h!]
\centering

% Left: TABLE
\begin{minipage}{0.48\textwidth} % keep under 0.5
  \captionsetup{type=table}
  \caption{HET WFU and HRS Parameters}
  \label{tab:parameters}
  \renewcommand{\arraystretch}{1.3}
  \begin{tabular}{|c|c|}
  \hline
  \textbf{HET Parameters} & \textbf{Value} \\
  \hline
  Diameter & 10.0 m \\
  F\# & 3.65 \\
  Image scale & 177 µm/'' \\
  ZDs & 35$^\circ \pm 6^\circ$ \\
  \hline
  \textbf{HRS Parameters} & \textbf{Value} \\
  \hline
  Wavelength range & 360 nm -- 1000 nm \\
  FOV & 35.6'' (6.3 mm) \\
  Fiber size & 1.47'' (260 µm), 1.81'' (320 µm) \\
  \hline
  \end{tabular}
\end{minipage}%
% \hfill
% Right: FIGURE
\begin{minipage}{0.48\textwidth}
  \centering
  \includegraphics[width=0.63\linewidth]{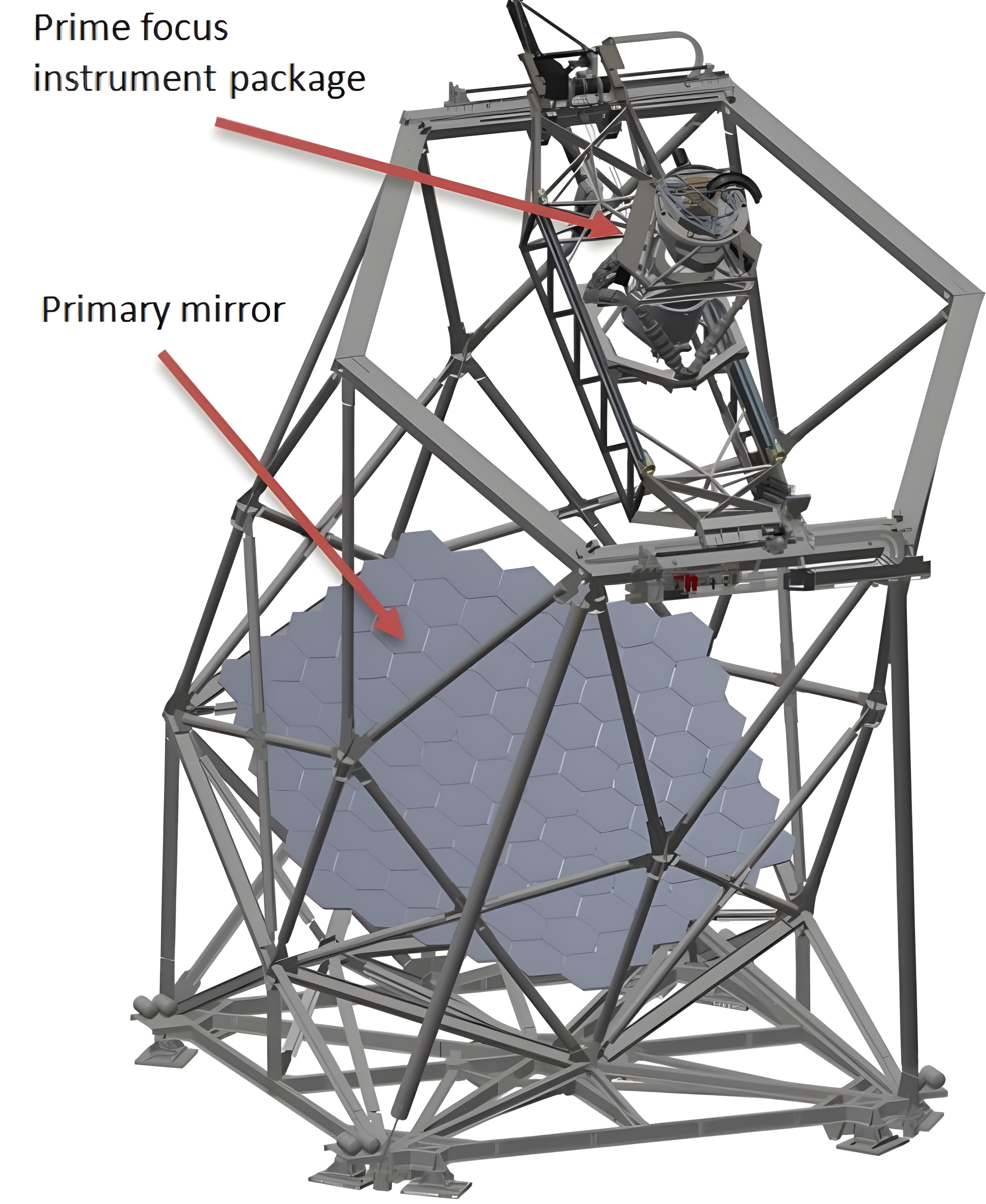}
  \captionsetup{type=figure}
  \caption{Hobby–Eberly Telescope structure.}
  \label{fig:het}
\end{minipage}

\end{figure}

The Hobby–Eberly Telescope (HET) is a 10 meter class optical telescope designed primarily for spectroscopy \cite{Hill2021}. Its High Resolution Spectrograph (HRS) was a first generation instrument that is being substantially upgraded for improved stellar and extragalactic spectroscopy. Building on the original HRS design \cite{Tull1998}, the upgraded HRS (sometimes referred to as the “HRS2”) is a single-object, fiber-fed echelle spectrograph equipped with dedicated sky fibers for simultaneous background subtraction. The spectrograph employs a white-pupil optical design (utilizing a two-pass collimator) with a mosaic of two R4 echelle gratings as the primary dispersing element. This configuration achieves broad wavelength coverage and high resolving power while keeping the spectrograph’s physical size manageable. A future upgrade will add a blue arm for continuous wavelength coverage between 362 and 680 nm. The existing arm has selectable Volume Phase Holographic (VPH) grating cross dispersers allowing coverage to 1000 nm. HRS has selectable resolving powers (R = 17,500 up to R = 105,000) by using different fiber and image slicer combinations. To balance light throughput against spectral resolution, the instrument integrates image slicers and offers two sizes of octagonal fiber cores ($\approx$ 1.47" and 1.81" on-sky diameters) for feeding the spectrograph. The larger fiber core diameter provides higher throughput for faint targets, whereas the smaller fiber yields higher spatial resolution and better background subtraction for targets in confused backgrounds. Figure \ref{fig:het} shows the rendered model of the 10 m HET, showing the fixed 35\textdegree elevation primary mirror and the tracking structure. The primary mirror consists of 91 hexagonal segments mounted on a steel truss structure. The tracker at the top end of the structure houses the Wide Field Corrector (WFC) and the Prime Focus Instrument Package (PFIP), which moves to track targets across the sky. The design allows a large collecting area at low cost, and is optimally used in a queue observing mode, making HET appropriate for spectroscopic surveys and follow-up observations. The key parameters of both the HET and the HRS fiber feed are summarized in Table \ref{tab:parameters}.

\noindent \textbf{Background \& Motivation}: Atmospheric dispersion is a well-known limitation in ground-based optical astronomy, and it is unfavorable to high-resolution spectroscopy. As starlight passes through Earth’s atmosphere, the wavelength-dependent refractive index of air causes differential refraction, displacing and elongating the stellar image along the dispersion axis. The magnitude of this effect increases with zenith distance (ZD) and can exceed the entrance aperture of a spectrograph’s fiber feed, leading to wavelength-dependent coupling losses. The effect is strongest at the blue end of the spectrum, resulting in a degraded signal-to-noise ratio (S/N), reduced radial velocity precision, and compromised continuum quality for blue wavelengths. This impact is significant for HET, because its fixed-altitude design, means that all observations occur at relatively large ZDs (high airmasses). In 2016, HET underwent a major Wide Field Upgrade (WFU) that expanded its field of view (FOV) by roughly an order of magnitude (to a diameter of $\sim$ 22 arcminutes) and improved its tracking capabilities \cite{Hill2018}. A key component of this upgrade was the installation of a new wide-field corrector (the Harold C. Simmons Optical System), which enabled HET to access a much larger field with reduced vignetting, improved image quality and the chief rays perpendicular to the focal surface at all field positions for fiber feeding. In the original configuration, the HRS fibers were fed directly at the telescope’s focal plane without any intermediate optics or atmospheric dispersion compensation. As a result, observations at the fixed 35\textdegree elevation suffered significant wavelength-dependent coupling losses, which affected high-resolution, blue-sensitive observations. For example, at a ZD of 35\textdegree\,(approximately 1.2 airmasses), uncorrected atmospheric dispersion can spread a stellar image by about 2.3" from 360 nm to 1000 nm. This blur exceeds the typical fiber core diameters (1.47" - 1.81"), causing throughput loss - especially in the blue.

\noindent To address this issue, the upgraded HRS fiber feed incorporates relay optics coupled with a static atmospheric dispersion corrector (ADC). %The ADC is implemented as a pair of Amici prisms that realign the dispersed light, effectively cancelling out the atmospheric differential refraction across the HRS’s full wavelength range. The relay optics re-image the telescope focal plane onto the fiber entrance, allowing the ADC-corrected beam to be efficiently coupled into the spectrograph fibers. 
By mitigating wavelength-dependent losses, this new fiber feed is expected to improve light injection efficiency, particularly at blue wavelengths, thereby enhancing the achievable S/N. This paper presents the design and implementation of the upgraded HRS fiber feed, with emphasis on the optical design and expected performance of the ADC, the resulting throughput improvements, the preliminary mechanical layout, and planned future work (including tolerance analysis and stray-light evaluation).

\section{OPTICAL DESIGN}

\begin{figure}[h!]
\centering

% Left: Table with caption above
\begin{minipage}[c]{0.45\textwidth}
  \captionsetup{type=table,position=top}
  \caption{Specifications of HRS ADC}
  \label{tab:adcparameters}
  \renewcommand{\arraystretch}{1.3}
  \begin{tabular}{|c|c|}
  \hline
  \textbf{Parameters} & \textbf{Value} \\
  \hline
  Glass Materials & \makecell[c]{BAL15Y ($n_{d}$ = 1.556, $v_{d}$ = 58.68)\\S-FPL51Y ($n_{d}$ = 1.497, $v_{d}$= 81.14)} \\
  Diameter & 11.6 mm \\
  Apex angles & 25.404\textdegree, 23.708\textdegree \\
  Center thicknesses & 4.55 mm , 4.39 mm \\
  Edge thicknesses\footnotemark & 1.99 mm, 7.28 mm, 7.12 mm, 1.5 mm \\
  \hline
\end{tabular}
\end{minipage}%
\hfill
% \hspace{5mm}
% Right: Figure with caption below
\begin{minipage}[r]{0.47\textwidth}
  % \hfill
  \centering
  \includegraphics[width=0.85\linewidth]{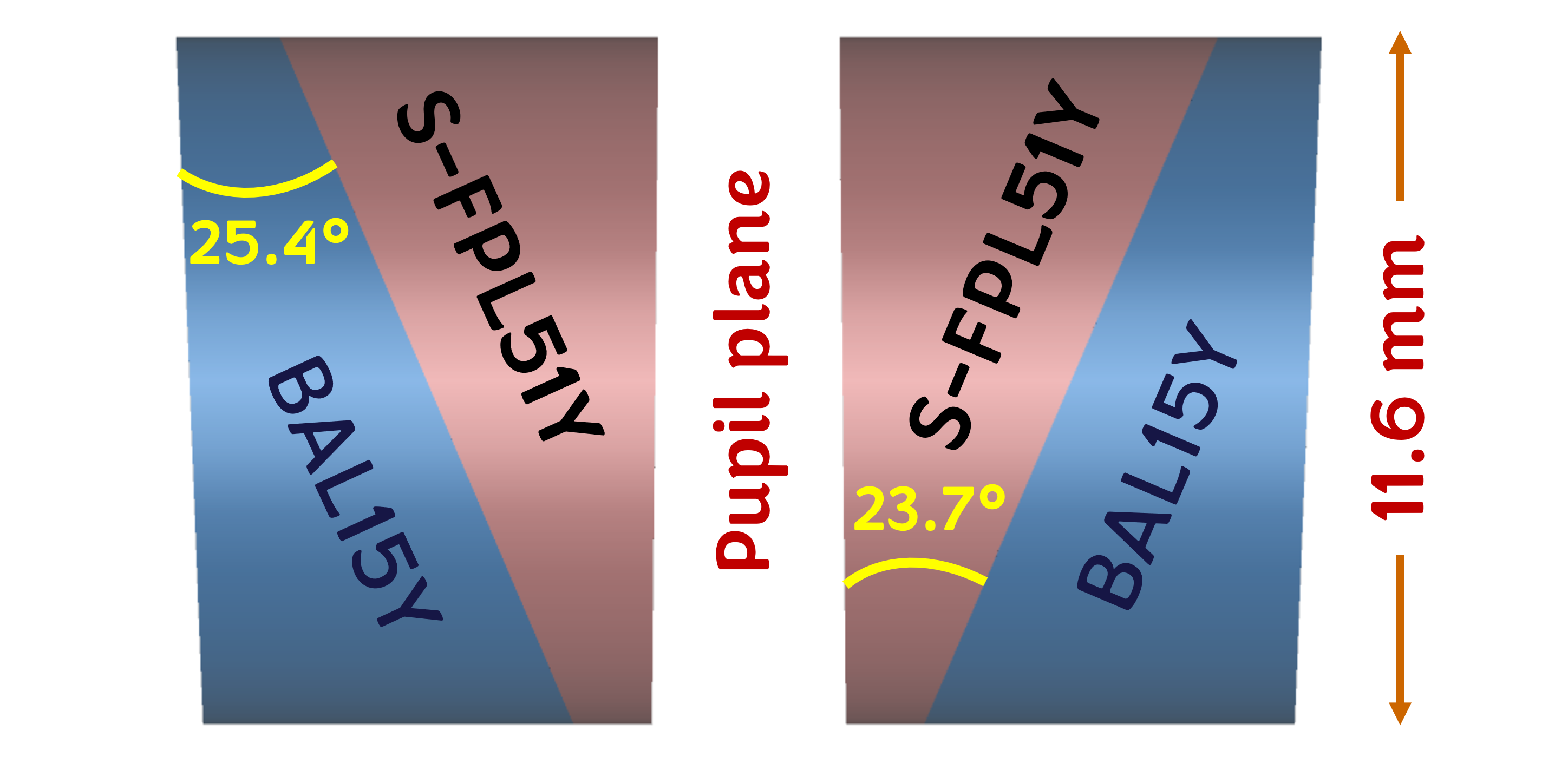} % replace with your actual image file
  \captionsetup{type=figure,position=bottom}
  \caption{Optical layout of ADC.}
  \label{fig:adc}
\end{minipage}

\end{figure}

The fiber feed for HRS is designed to deliver a seeing-limited target image from the HET focal surface (FS) to the spectrograph with minimal loss or aberration. In the original HRS design, f/4.6 light from the HET was coupled the science fiber (and a separate fiber for the sky). With an f/3.65 HET beam, relay optics that incorporate an ADC are currently being incorporated to counteract atmospheric refraction. The most efficient and widely adopted approach for atmospheric dispersion correction in high-resolution spectrographs is the Rotating Atmospheric Dispersion Corrector (RADC)\cite{Bestha2023,Born2022}. Additionally, the volume constraints of the input head, combined with the resulting lateral image shift, rendered a Linear ADC design impractical. We started the optical design based on counter-rotating Amici prisms\cite{Wynne1984}. The RADC design uses two identical Amici prisms, each producing no beam deviation at the design wavelength, mounted to rotate about the optical axis. With their dispersion axes aligned, the prism effects combine to produce maximum dispersion. When rotated 180\textdegree relative to each other, the dispersion is completely canceled. Any intermediate rotation angle provides partial correction, allowing adjustment for different ZDs. A high-performance input head with an active ADC was initially designed; however, the HET program had concerns about its long-term reliability leading us to investigate the feasibility of a static ADC. Preliminary analysis itself indicated that, over the limited range of HET ZDs (26.5\textdegree\,to 43.5\textdegree), a dynamic ADC would offer some performance gains compared to a static design. The static design does not allow the ADC dispersion axis to be aligned parallactic over a full track. In the worst case (when the HET azimuth is zero), the error will range from -13.7\textdegree\,to +13.7\textdegree. The current ADC is a static design comprising two identical Amici prisms, fixed at an orientation optimized for a ZD of 35\textdegree. This configuration effectively corrects atmospheric dispersion over a ZD range of 26.5\textdegree\ to 43.5\textdegree. By setting appropriate glass materials and apex angles, the prisms introduce a dispersion of opposite sign to the atmosphere’s dispersion, so the net dispersion is nearly zero. Atmospheric dispersion for different ZDs at the HET site was simulated using the ZEMAX atmospheric model surface\cite{}, incorporating site-specific parameters such as altitude, humidity, pressure, and temperature. The ADC in our design corrects dispersion across the full operational band (360 - 1000 nm). The initial material selection considered all optical glasses in the ZEMAX glass catalog with transmission exceeding 99\% for a 10 mm thickness at 360 nm. This set was later narrowed to i-line glasses from OHARA\footnote{\url{https://www.oharacorp.com/product-category/i-line/}} and NIKON\footnote{\url{https://www.nikon.com/business/components/lineup/materials/i-line/}}, chosen for their high homogeneity, excellent transmission characteristics, and availability. The specifications of the ADC are given in Table \ref{tab:adcparameters}. Without the ADC, blue and red wavelengths focus at positions separated by $\sim$ 2.3" at a ZD = 35\textdegree, whereas with the ADC the residual dispersion is reduced to less than one-quarter of the 1.47" fiber diameter at the extreme HET ZDs. 

\footnotetext{The values are measured at the top and bottom edges of the first prism set, as shown in the ADC design in Figure \ref{fig:adc}.}

\begin{figure}[h!]
  \centering
  \includegraphics[width=\textwidth]{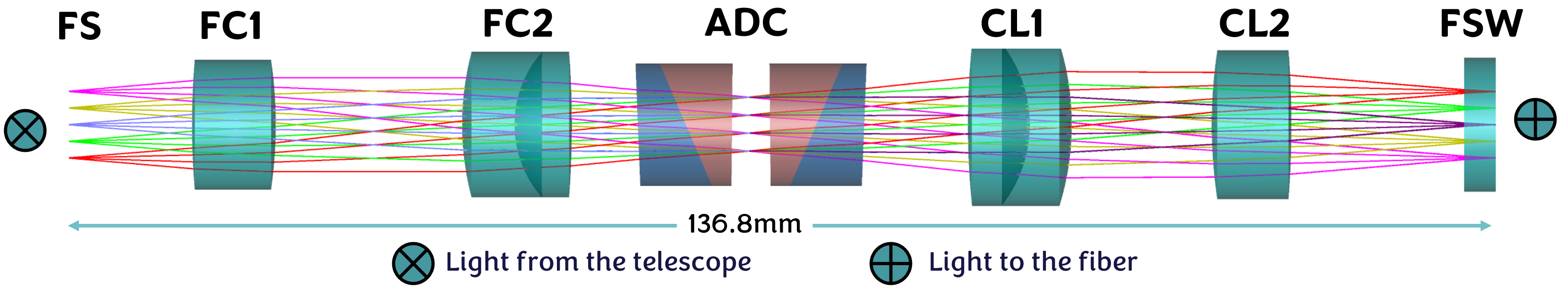} % make sure this file is uploaded to Overleaf
  \caption{
  Optical layout of the HRS fiber feed system. Light from the HET enters from the left and is focused onto the fiber at the right. The system includes the following labeled components:
  {FS}: HET Focal Surface, 
  {FC1}: Fabry Collimator 1 (BSL7Y), 
  {FC2}: Fabry Collimator 2 (PBL6Y + S-FPL51Y), 
  {ADC}: Atmospheric Dispersion Corrector (BAL15Y + S-FPL51Y), 
  {CL1}: Camera Lens 1 (S-FPL51Y + PBL25Y), 
  {CL2}: Camera Lens 2 (S-FPL51Y), 
  {FSW}: Fused Silica Window.
  }
  \label{fig:fiberfeed}
\end{figure}

In Figure~\ref{fig:fiberfeed}, light from the HET focal surface (FS) propagates through a sequence of optical elements before being coupled into the fiber on the right. Fabry Collimator 1 (FC1) controls the beam diameter and, along with the Fabry Collimator 2 (FC2) doublet, collimates the light for downstream optics and forms a pupil in between the fixed ADC. The ADC comprises of two identical Amici prism assemblies with similar apex angles and glass materials; the second prism set is rotated by 180\textdegree\ about the vertical (Y) axis to cancel atmospheric dispersion at the operating wavelength and nominal ZD. Following dispersion correction, the Camera Lenses (CL1 and CL2) refocus the beam onto the fiber input. CL1 is a doublet (S-FPL51Y + PBL25Y) and CL2 is a singlet (S-FPL51Y), with both designed to minimize aberrations and deliver a well-corrected, sharply focused, telecentric image at the fiber input. A Fused Silica Window (FSW) is placed to be sufficienlty out of focus to minimize effects of dust, surface, and coating blemishes.

\section{PERFORMANCE ANALYSIS}
\label{sec:sections}

\begin{figure}[h!]
  \centering
  \includegraphics[width=0.67\textwidth]{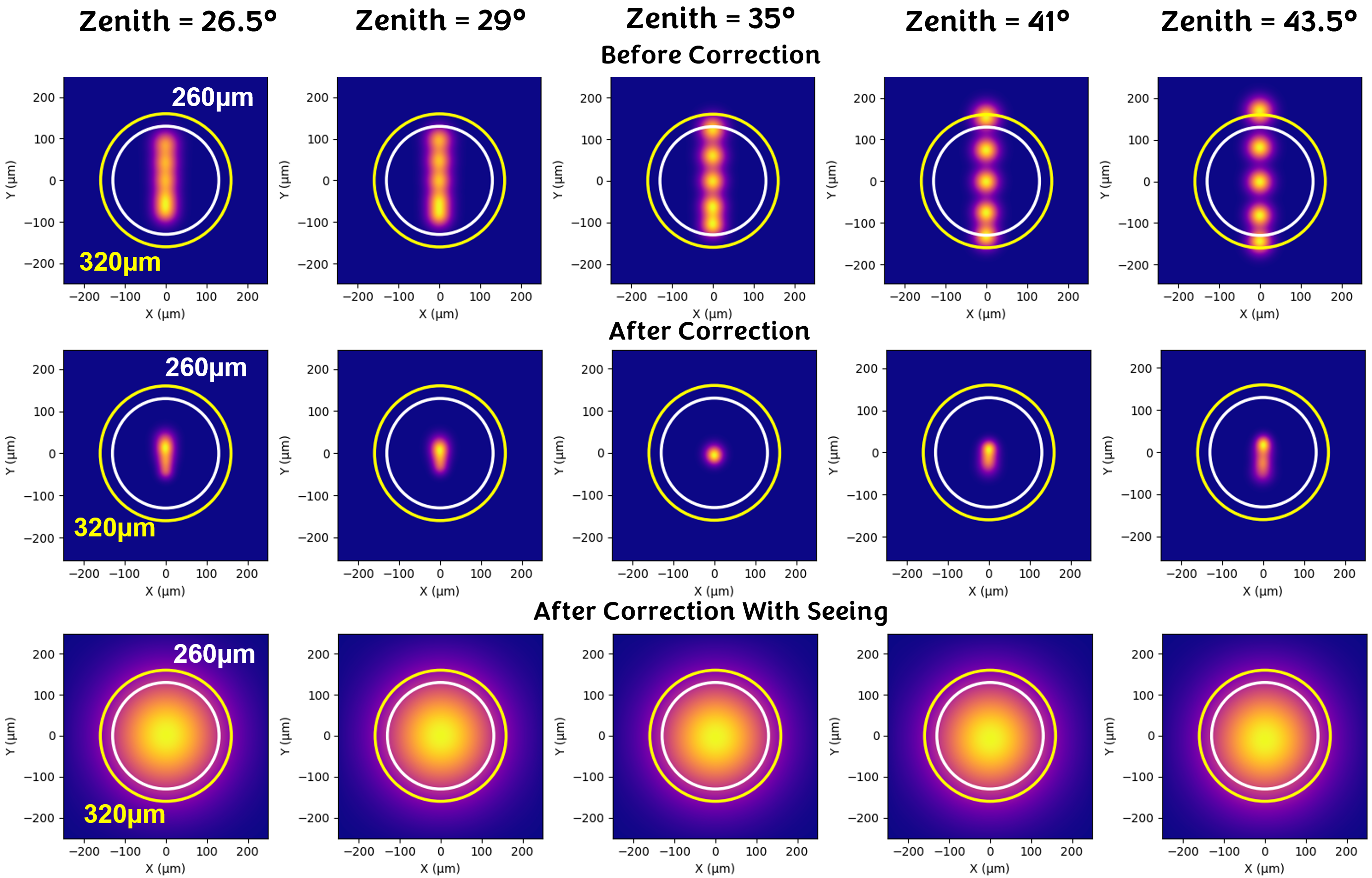}
  \caption{Simulated PSFs at five wavelengths between 360 and 1000 nm, and various ZDs (26.5\textdegree to 43.5\textdegree) for the on-axis field (0",0"). 
  \textbf{Top row:} PSFs before atmospheric dispersion correction, showing strong chromatic elongation.  
  \textbf{Middle row:} PSFs after correction using the fixed ADC, demonstrating significantly improved spatial confinement across wavelengths.  
  \textbf{Bottom row:} PSFs after correction including the effect of atmospheric seeing, showing realistic beam profiles. White and yellow circles represent 260 µm (1.47") and 320 µm (1.81") fiber core diameters.
  }
  \label{fig:psf_seeing}
\end{figure}

\textbf{Atmospheric Dispersion Correction}: A primary result of this design upgrade is the effective correction of atmospheric dispersion by the ADC. Figure \ref{fig:psf_seeing} illustrates the improvement in dispersion. In the top panel, we plot the simulated point spread functions (PSFs) at the HET focal surface at wavelengths from 360 nm to 1000 nm for the range of ZDs $\approx$ 26.5\textdegree - 43.5\textdegree\,without the ADC. PSFs were generated by considering the Root Mean Square (RMS) spot diameter from the ZEMAX model as the full width at half maximum (FWHM) of the PSF. The RMS spot radii for all wavelengths and ZDs were obtained from ZEMAX model using the ZEMAX-Python interface package, PyZDDE\footnote{\url{https://pypi.org/project/PyZDDE/}}.The simulations were conducted only for the on-axis field. The top row in Figure \ref{fig:psf_seeing} shows the dispersion of light for different ZDs blue light (top) is offset by a few arcseconds from red light (bottom), and intermediate wavelengths spread out along a roughly vertical line since atmospheric dispersion occurs largely in the altitude direction. This would cause significant losses with a fiber of $\sim$ 1.47" – 1.81" diameter. Quantitatively, our simulation shows about 2.3" of dispersion between 360 nm and 1000 nm at ZD = 35\textdegree. The middle panel of Figure \ref{fig:psf_seeing} shows the same simulation with the fixed ADC incorporated: the multi-wavelength PSFs now almost overlap within $\leq$ 0.42" for the extreme ZDs (i.e. quarter of the 1.47" fiber size for the extreme ZDs), ensuring that essentially all wavelengths fully enter the fiber together. In terms of throughput, without an ADC the fiber would only transmit efficiently a narrow wavelength band centered near the guide wavelength (e.g. $\sim$ 462 nm), whereas with the ADC we achieve high coupling efficiency across the entire wavelength range. We estimate that $\geq$ 90\% of the light from 360 - 1000 nm now falls into the fiber for ZD 26.5\textdegree - 43.5\textdegree\,when using the ADC. The spot size at the fiber input is dominated by the input seeing (the HET specification is 1.3") convolved with small residual aberrations. In the bottom panel, the atmospheric dispersion corrected spots are convolved with a typical HET seeing PSF having a FWHM of 1.3". 

\begin{figure}[h!]
  \centering
  \includegraphics[width=0.55\textwidth, trim = {0cm 0cm 0cm 0.5cm}, clip]{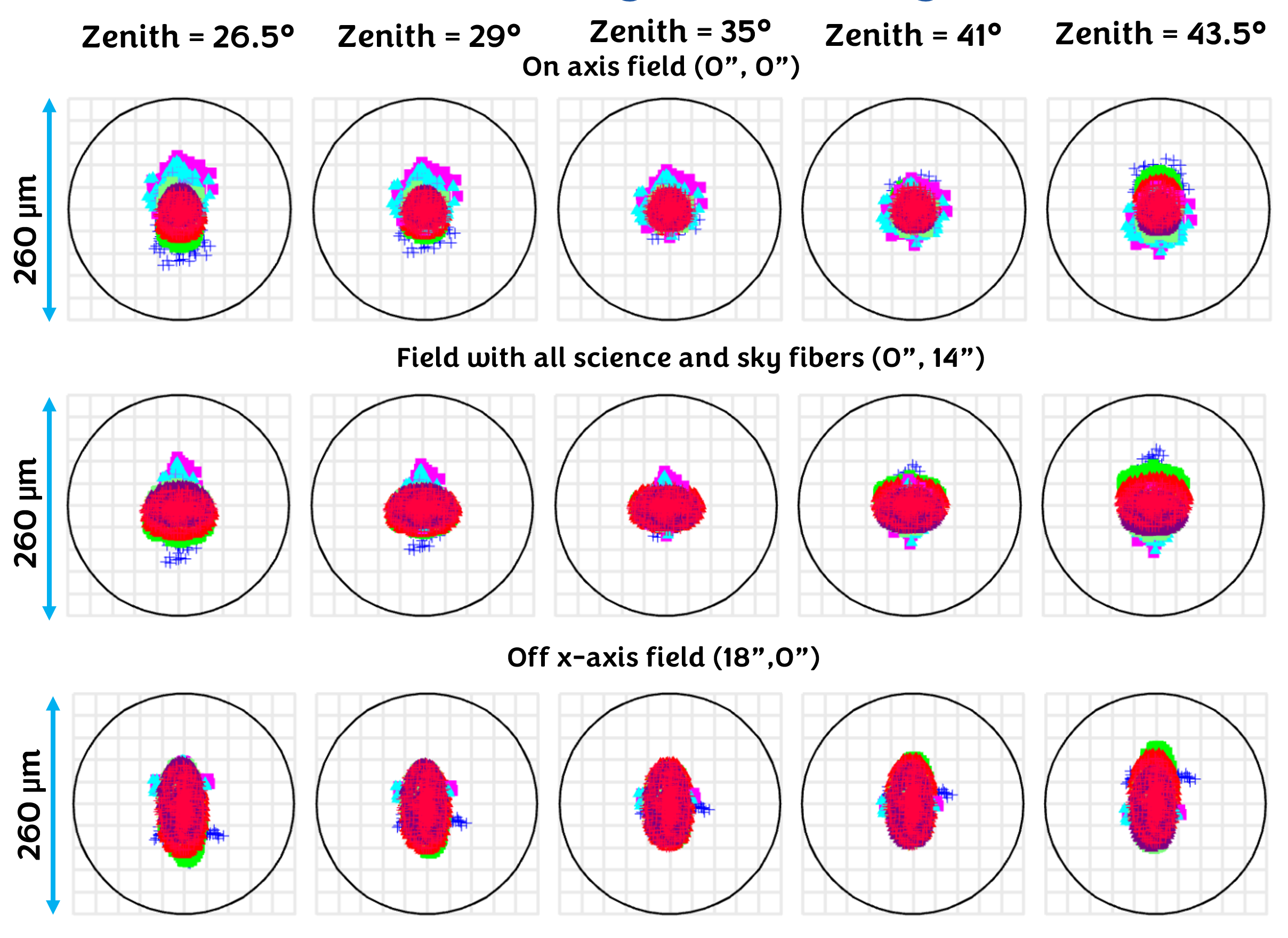} % make sure this file is uploaded to Overleaf
  \caption{Spot diagrams after atmospheric dispersion correction for the full range of HET ZDs (26.5\textdegree to 43.5\textdegree), across three different field positions. Each sub-panel shows the geometric distribution of light from 360 nm to 1000 nm, with colors representing different wavelengths. \textbf{Top row:} On-axis field (0", 0"). \textbf{Middle row:} Intermediate field position which bound the region occupied by all science and sky fibers (0", 14"). \textbf{Bottom row:} Extreme off-axis field along the $x$-direction (18", 0"). The black circle indicates the fiber core diameter (260 µm).}
  \label{fig:imagequality}
\end{figure}

\noindent \textbf{Image Quality}: In addition to evaluating dispersion, we assessed the image quality at the fiber input using the RMS spot radius obtained from the ZEMAX spot diagram analysis. The encircled energy diameter at the fiber input is $\sim$ 0.22" (80\% EE) at ZD = 35\textdegree \ for the on-axis field, and $\sim$ 0.32" for the extreme FOV. We note that the science fibers are in the inner field and the sky fibers are in the mid field. Figure~\ref{fig:imagequality} presents polychromatic spot diagrams for the static ADC design across the operational ZD range of 26.5\textdegree - 43.5\textdegree. The results show that, even with a static ADC, the system delivers consistently comparable image quality. For all wavelengths and ZDs considered, the RMS spot radii remain well within the 260 µm fiber core diameter indicated in the diagrams. The relay optics not only maintain image quality across the zenith range but also deliver performance comparable to the HET at the nominal ZD of $\sim$ 35\textdegree. Across the full FOV, the spot diagrams show compact and well-confined spots, including the on-axis position (0", 0"), intermediate positions such as (0", 14"), which correspond to locations occupied by all science and sky fibers, and extreme $x$-axis positions (18", 0"). These results confirm that the static ADC and relay optics combination preserves image quality over the entire science field.

\begin{figure}[h!]
  \centering
  \includegraphics[width=0.73\textwidth, trim={0cm 0cm 0cm 3.5cm}, clip]{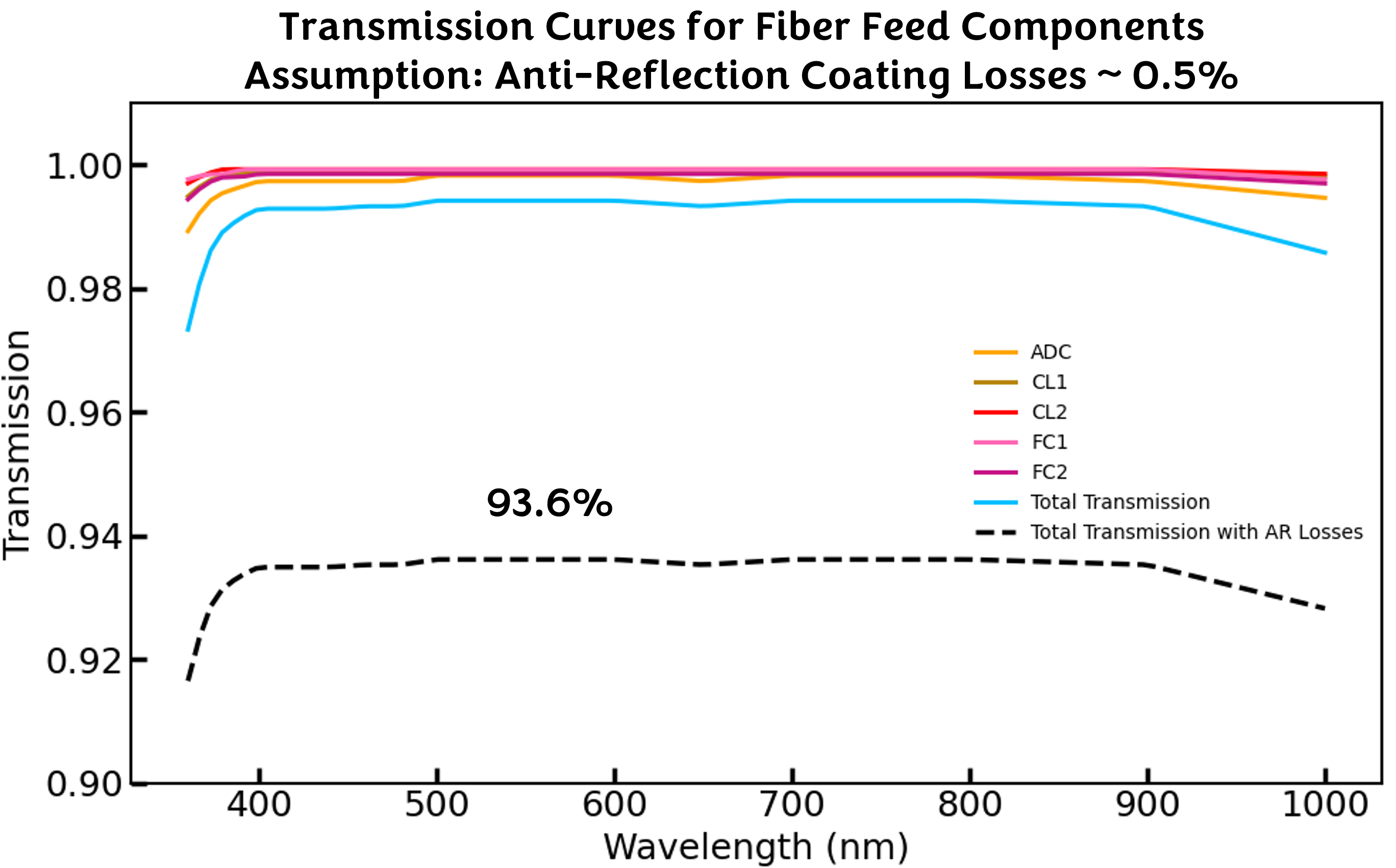}
  \caption{
  Transmission curves for individual fiber feed components and the overall system. The solid lines represent the transmission of each component: ADC, Camera Lens 1 (CL1), Camera Lens 2 (CL2), Fabry Collimator 1 (FC1), and Fabry Collimator 2 (FC2). The total optical transmission (excluding AR losses) is shown in blue ($\ge$ 97.3\%). The dashed black line includes an estimated 0.5\% loss per surface due to AR coating performance, resulting in a throughput $>$ 91.6\%, with a typical value of 93.6\%.  We are also pursuing higher performance AR coatings with less than 0.1\% loss per surface.
  }
  \label{fig:transmission}
\end{figure}

\noindent \textbf{Optical Transmission and Light Loss at Fiber Input}: To evaluate the efficiency of ADC-based HRS fiber feed optics, we analyzed both the internal and external optical transmission of individual components and the overall light injection efficiency into the science fibers (1.47" and 1.81"). Figure \ref{fig:transmission} shows the wavelength-dependent transmission of each optical element - ADC, collimators, and camera lenses - modeled using ZEMAX Internal Transmission Analysis (ITA) for the selected glass materials. The different colored curves represent the internal transmission corresponding to the thickness of each optical element, with the largest value between the center and edge thicknesses used in the calculation. For example, for the first lens in the FC2 doublet, the edge thickness exceeds the center thickness; therefore, the edge thickness value is used to determine the internal transmission in ZEMAX ITA. Taking into account typical astronomical anti-reflection (AR) coatings\footnote{\url{https://spectrumthinfilms.com/stf/astronomical-coatings-optics/}} with $\approx$ 0.5\% loss per surface, the total transmission throughput of the system is estimated to be $>$ 91.6\% in the 360 - 1000 nm wavelength band. The fiber feed optical design achieves an overall transmission throughput of $\geq$ 91.6\%, with a typical value of 93.6\%.

\begin{figure}[h!]
  \centering
  \includegraphics[width=0.75\textwidth, trim={0cm 0cm 0cm 2.75cm}, clip]{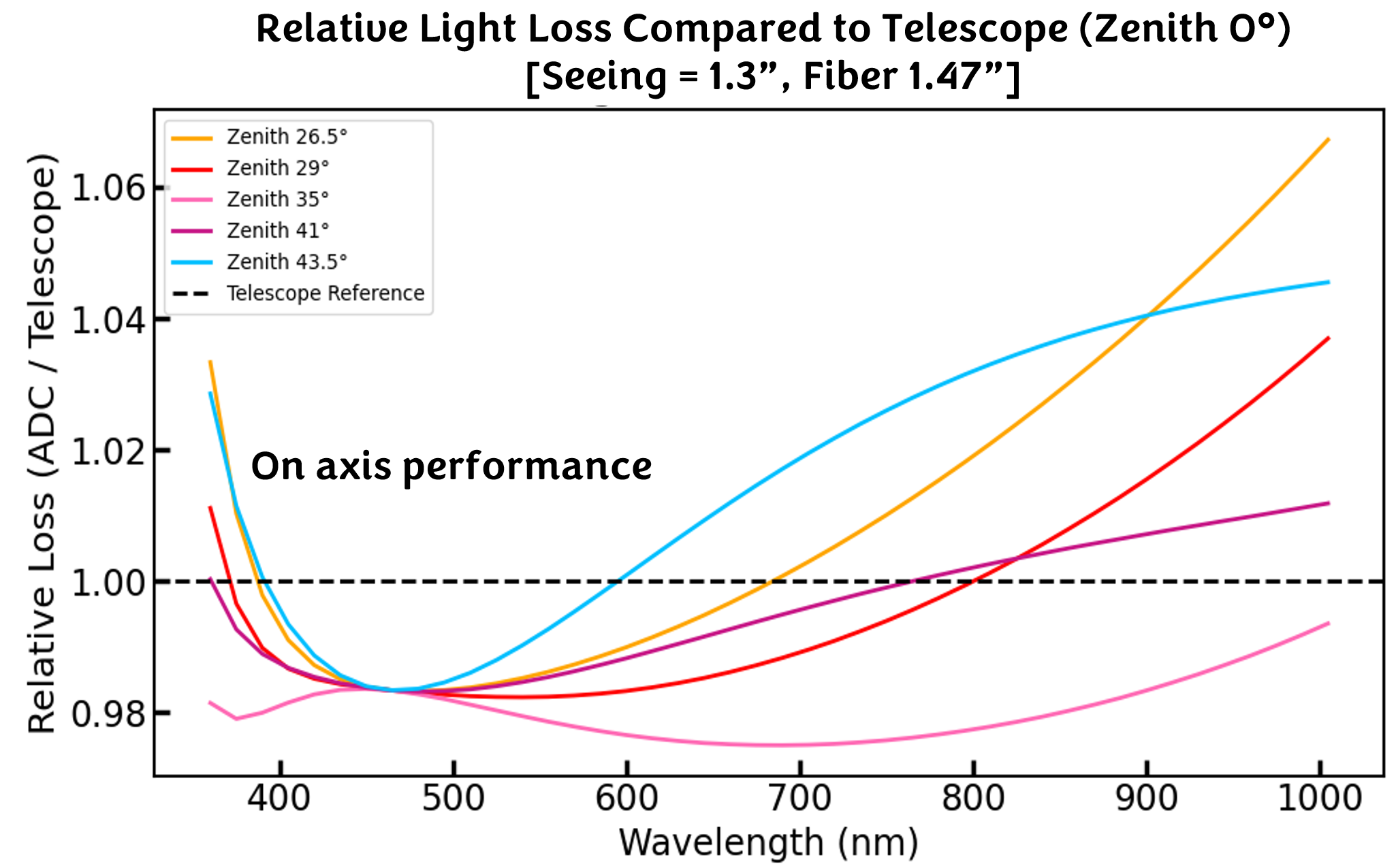}
  \caption{Relative light loss of the ADC - based HRS fiber feed optics compared to an HET - only case at ZD = 0\textdegree, for the HET ZD range of 26.5\textdegree\,to 43.5\textdegree. Simulations assume 1.3" seeing FWHM and a 1.47" fiber diameter. The colored curves show the wavelength-dependent injection losses normalized to the HET PSF without atmospheric dispersion. The dashed black line represents the HET-only reference case. %At all ZDs, the system maintains $\leq$ 6.8\% additional loss (for extreme ZDs and wavelengths), with the least loss in the 400 - 600 nm range.
  In the worst combination of wavelength, ZD, and tracker position, the system comes within 6.8\% of the HET's atmospheric dispersion free performance. The system slightly outperforms the bare HET at prime wavelengths and ZD.}
  \label{fig:light_loss}
\end{figure}
We also quantified the relative light injection efficiency by comparing the relayed and corrected PSFs at various ZDs to a dispersion-free HET reference at ZD = 0\textdegree. The PyZDDE Python package was used to extract the RMS spot radius by iteratively varying the field position, wavelength, and ZD in the ZEMAX model. The RMS spot radius was used to generate the PSF, which was then convolved with a seeing profile having a FWHM of 1.3". Light loss was computed geometrically by masking the convolved PSF with the science fiber aperture (260 µm or 320 µm). Figure~\ref{fig:light_loss} shows the wavelength-dependent light loss for particular ZDs, incorporating the seeing conditions (1.3") and the smallest fiber diameter (1.47"). The results indicate that the ADC design successfully minimizes wavelength-dependent loss. Across the full wavelength range (360 - 1000 nm), the loss introduced by the ADC is minimal, ranging from approximately -2\% to 6.8\% depending on ZD and wavelengths, with the largest losses occurring at the extremes of the spectral band for the highest ZDs. At ZD = 35\textdegree\,(pink curve), there is a net gain of slightly under 2\%, which is as expected the best overall performance among the cases. The fiber input head relay optics correct some of the HET WFC residual aberrations. For nearly all ZDs, the minimum loss occurs around 400–600 nm, where the curves approach or fall below the HET reference line, corresponding to a relative gain of about 1.0 or less. The fiber feed optical design provides improved light injection efficiency at the blue end of the wavelength band. The telescope pointing system can be utilized to center a specific wavelength on the fiber core, enhancing image quality and light injection efficiency at the desired wavelength (centers Near-InfraRed (NIR) wavelengths on the fiber core when NIR wavelengths are the primary science).
\section{MECHANICAL DESIGN}
Designing the mechanical housing and positioning system for the fiber feed is a critical aspect of this project. The mechanical design must ensure optical stability and precise alignment of all components - particularly the ADC prisms under varying observing conditions (gravity orientations, temperature fluctuations, etc.) on the HET’s moving tracker. The fiber input head (IH), which houses the collimator and camera lens assembly, the ADC, and the fiber puck, is mounted on the Input Head Mount Plate (IHMP) located within the HET’s Prime Focus Instrument Package (PFIP, see Figure \ref{fig:het}) on the HET\cite{Vattiat}. The PFIP is mounted on the tracker and moves with the telescope; therefore, stiffness and volume constraints were important considerations. For example, as  the HET tracker moves in six degrees of freedom to follow targets, the mechanical design must ensure that the IH/fiber does not undergo significant flexure. As the tension on the fiber conduit changes to accommodate tracker motion, special care must be taken to avoid tight bends in the fiber cables, which helps preserve the focal ratio and prevent added FRD. The mechanical design, therefore, should provide a robust platform for the optical elements with minimal maintenance. The careful attention to alignment, materials, and mounting will ensure long-term stability of the fiber feed system once it is commissioned on HET. 
\begin{figure}[ht!]
  \centering
  \begin{minipage}[c]{0.48\textwidth}
    \centering
    \includegraphics[width=0.69\textwidth]{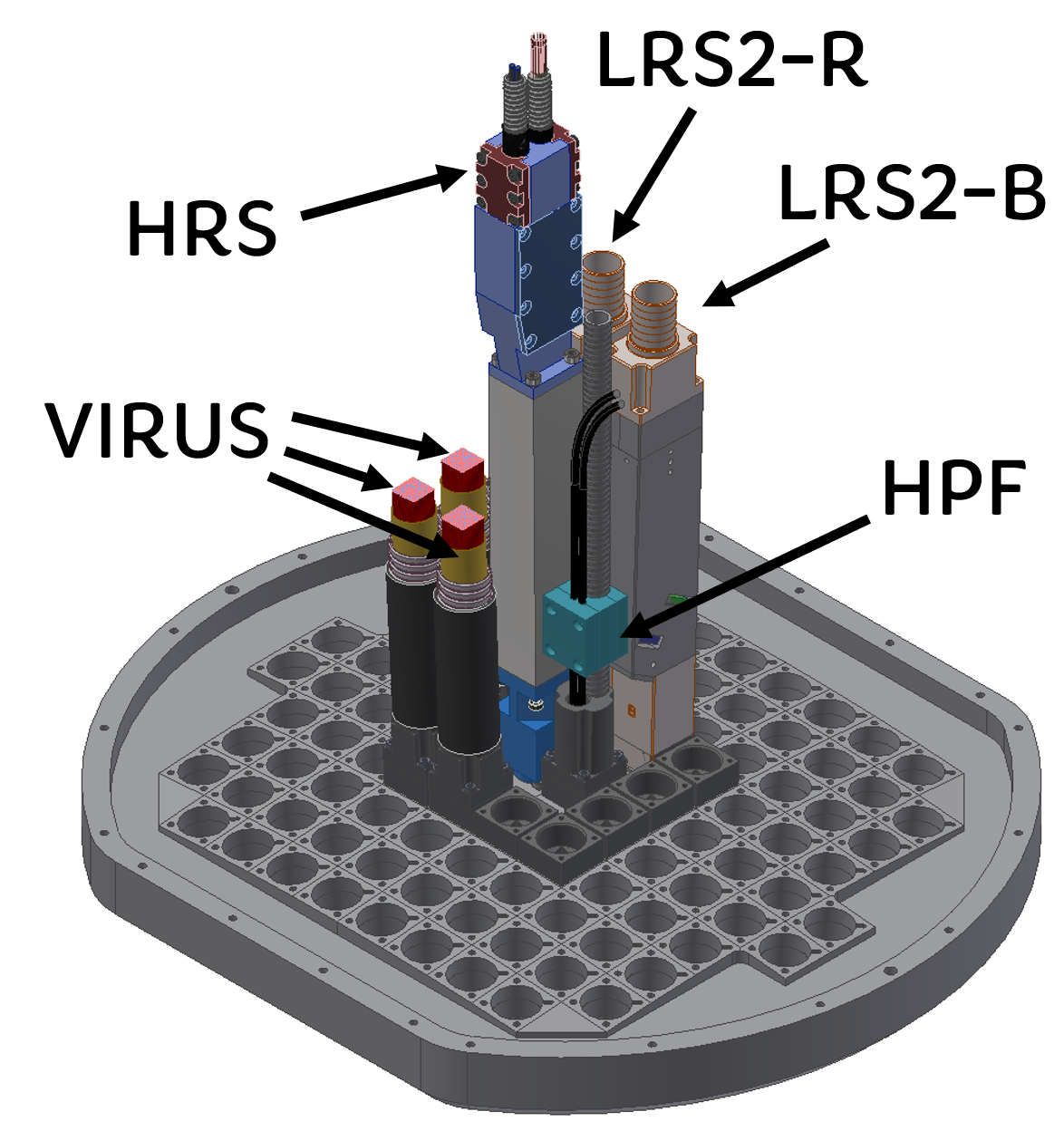}
    \caption{CAD rendering of the Input Head Mount Plate (IHMP) at the HET corrector focal surface, showing the fiber input heads of all HET instruments. The labeled components include VIRUS (only three IHs of 78 VIRUS IHs are shown for clarity), HPF, LRS2-B, LRS2-R, and the current HRS input head.}
    \label{fig:ihmp}
  \end{minipage}
  \hfill
  \begin{minipage}[c]{0.48\textwidth}
    \centering
    \includegraphics[width=0.5\textwidth]{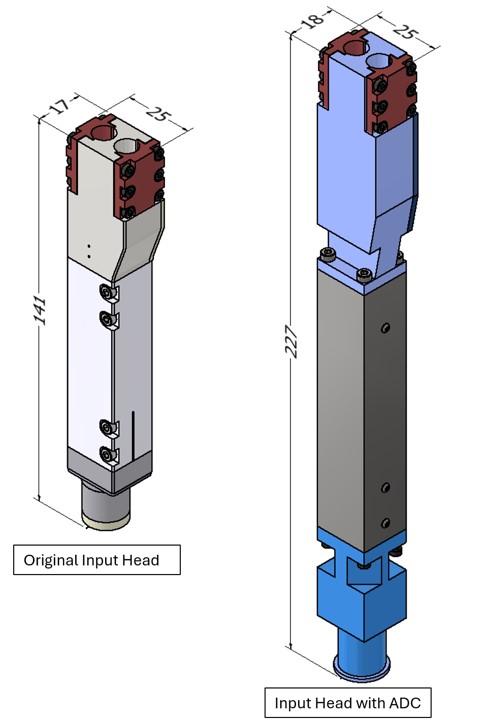}
    \caption{Comparison of dimensions between HRS Input Heads (IH) with and without an ADC, showing the $\sim$ 61\% increase in length to accommodate the static ADC and relay optics. Dimensions are in mm. There is not much change in width and height of original and redesigned IH. }
    \label{fig:ih_comp}
  \end{minipage}
\end{figure}
\vspace{-0.5cm}

\noindent A CAD rendering of the IHMP is shown in Figure \ref{fig:ihmp}, where the fiber input heads of HRS and other HET instruments are visible. In its current configuration, the HRS Input Head is the longest of all instrument input heads on the IHMP. As shown in Figure~\ref{fig:ih_comp}, the redesigned HRS IH, incorporating the static ADC and relay optics, is 227 mm long, corresponds to a $\sim$ 61\% increase over the original 141 mm design, while maintaining the same width and height to accommodate the additional optical components.

\begin{figure}[h!]
  \centering
  \includegraphics[width=0.98\textwidth, trim = {0mm 0cm 0mm 0cm}, clip]{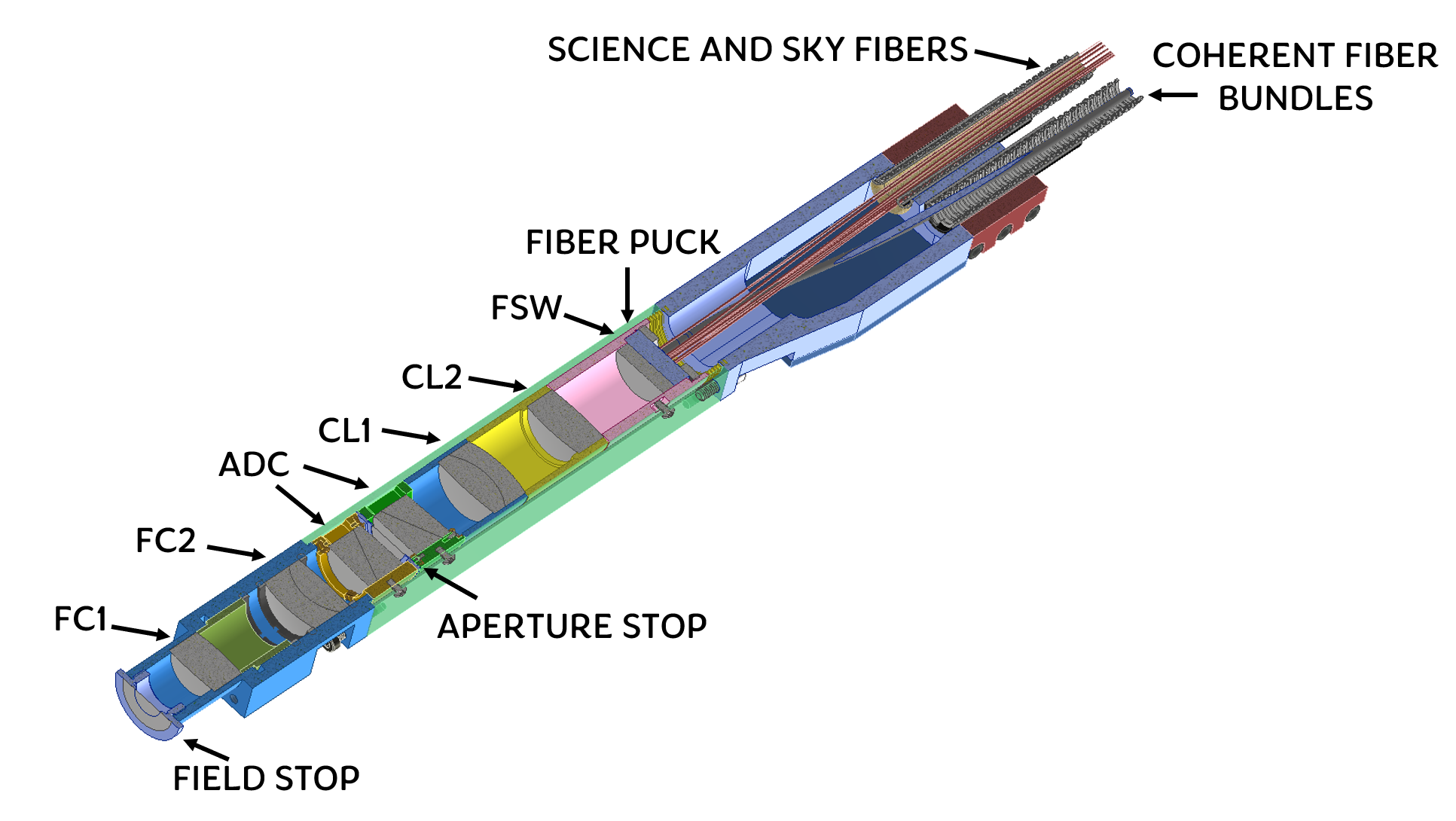} % Replace with your uploaded file name
  \caption{
  Cross-section of the HRS Input Head (IH) assembly. The design consists of three main stainless steel subassemblies: (1) collimator lenses (FC1 and FC2), (2) the dispersion correction and imaging optics including the ADC, camera lenses (CL1 and CL2), fused silica window (FSW), and fiber puck, and (3) fiber clamps for routing science, sky, and coherent fiber bundles. This modular configuration supports precise alignment, mechanical stability, and ease of integration into the Input Head Mount Plate (IHMP).
  }
  \label{fig:hrs_cross_section}
\end{figure}

A conceptual mechanical layout has been developed to support and align the optical components of the HRS fiber feed. The fiber input head assembly integrates the complete optical train and fiber interface into a mechanically rigid housing, ensuring precise alignment and long-term stability under operational conditions. The mechanical design of the HRS input head and feed assembly, shown in Figure \ref{fig:hrs_cross_section}, consists of three stainless steel subassemblies: the first houses the collimator lenses (FC1 and FC2); the second contains dispersion correction and imaging optics, including the ADC, camera lenses (CL1 and CL2), fused silica window (FSW), and fiber puck; and the third positions the fiber clamps for routing the science, sky and coherent fiber bundles. This modular configuration provides precise alignment, ease of integration into the IHMP, and facilitates future servicing.

The incoming telescope beam enters through a field stop, which defines the beam acceptance, blocks unwanted field rays, and aids in baffling against scattered light. Each element is separated by spacers that set the exact optical air gaps defined in the ZEMAX prescription. This ensures that lens positions along the optical axis remain within tight tolerances. Retainers secure the optics axially, preventing displacement due to gravity vector changes, or telescope tracking motion, while avoiding stress-induced aberrations. The prisms, being the most sensitive optical components, are mounted in precision carriers (gold and dark green annular parts) and fastened by radial screws that prevent shifting under gravity or vibration, locking them in place after alignment. An aperture stop is positioned at the system’s pupil to control stray light, maintain the designed f-number, and limit the beam size. At the output, the fiber puck precisely holds and aligns the science fibers, sky fibers, and coherent fiber bundles at the telecentric focal plane, ensuring stable, repeatable coupling to the spectrograph pre-slit optics.

\vspace{-0.5cm}
\section{CONCLUSION AND Future Work}
We have developed an input fiber feed for the HET's HRS that enhances its performance for high resolution spectroscopy. This upgrade increases throughput and blue-end sensitivity, enabling higher S/N, more precise radial velocity measurements, and improved continuum accuracy in HRS spectra - particularly at blue wavelengths. Key achievements and plans are summarized below:

\noindent\textbf{Design Accomplishments}: 
A static ADC employing Ohara i-line glasses achieves $>$ 91.6\% transmission over the 360 - 1000 nm bandpass and effectively corrects atmospheric dispersion for zenith angles from 26.5\textdegree\,to 43.5\textdegree. Simulations indicate that over the full range of HET ZDs, atmospheric dispersion is well corrected (residual $<$ 0.42"). The relay optics improve light injection efficiency - particularly at blue wavelengths - enhancing sensitivity for high-resolution spectroscopy. The design demonstrates improved throughput across the full operating bandpass and maintains high image quality across the entire field of view. The co-optimized relay optics improve fiber coupling efficiency, particularly at the blue end, and yield image quality better than the HET-delivered image quality at the nominal ZD of $\sim$ 35\textdegree. 

\noindent\textbf{Innovations}: This work demonstrates that instead of a dynamic/rotating ADC, a static ADC, optimized for the HET’s fixed-altitude design, can deliver high-performance atmospheric dispersion correction by carefully selecting high-transmission glasses and refining prism apex angles for HET's specific zenith range (26.5\textdegree–43.5\textdegree).

\noindent\textbf{Future Work}: The optical and mechanical design will be further refined through detailed tolerancing and ghost analysis to ensure manufacturability, alignment stability, and stray light control.

In conclusion, the redesigned HRS fiber feed optics for HET has been successfully developed to meet the requirements of high throughput and image quality. By combining the optical design with robust mechanical design, we are in the process of implementing a system that significantly improves HET-HRS throughput. This upgrade will enable the acquisition of higher-quality spectra and ensure that HET remains a leading facility for precision spectroscopic science.

\acknowledgments % equivalent to \section*{ACKNOWLEDGMENTS}   
Devika Divakar thanks Manjunath Bestha for the technical discussions.  The Hobby-Eberly Telescope (HET) is a 10 meter class telescope run by an international collaboration among The University of Texas at Austin, The Pennsylvania State University, Ludwig-Maximilians-Universität München, and Georg-August-Universität Göttingen. The HET is named in honor of its principal benefactors, Lt. Governor William P. Hobby, Jr., of Texas and Robert E. Eberly of Pennsylvania. 
% References
\bibliography{report} % bibliography data in report.bib
\bibliographystyle{spiebib} % makes bibtex use spiebib.bst
% R. G. Tull, “High-resolution fiber-coupled spectrograph of the Hobby-Eberly Telescope,” Proc. SPIE 3355, 387 (1998). (Original HET HRS design overview)

% HET Wide Field Upgrade Team, “Deployment of the Hobby-Eberly Telescope wide-field upgrade,” Proc. SPIE 10700, 1070013 (2018). (HET wide-field upgrade increasing pupil to 10 m and field to 22′)

% D. G. Bramall et al., “The SALT HRS spectrograph: final design, instrument capabilities and modes,” Proc. SPIE 7735, 77354F (2010). (Comparable fiber-fed high-res spectrograph with ADC on SALT)

\end{document}